# A Survey of Prediction Using Social Media

Sheng Yu and Subhash Kak

Department of Computer Science, Oklahoma State University

Stillwater, Oklahoma, U.S.A. 74078

{yshe, subhashk}@cs.okstate.edu

**Abstract:** Social media comprises interactive applications and platforms for creating, sharing and exchange of user-generated contents. The past ten years have brought huge growth in social media, especially online social networking services, and it is changing our ways to organize and communicate. It aggregates opinions and feelings of diverse groups of people at low cost. Mining the attributes and contents of social media gives us an opportunity to discover social structure characteristics, analyze action patterns qualitatively and quantitatively, and sometimes the ability to predict future human related events. In this paper, we firstly discuss the realms which can be predicted with current social media, then overview available predictors and techniques of prediction, and finally discuss challenges and possible future directions.

## 1. Introduction

Social media are platforms that allow common persons to create and publish contents. Two worldwide popular social media websites, Twitter and Facebook, demonstrate its explosive growth and profound influence. Both Twitter and Facebook are in the top 10 most-visited websites in the world according to Alexa ranking [1]. Facebook has more than 800 million active users [2], and by March 2011, on Twitter, there were about 140 million information pieces created and transferred daily [3]. There is other specialized social media that are focused on entertainment, sports, finance and politics.

Since there are many users sharing their opinions and experiences via social media, there is aggregation of personal wisdom and different viewpoints. Such aggregation has limitations as viewpoints are subject to change with time. In a sense the social media prediction problem is paralleled by prediction of financial time series based on past history, which has its uses in trading. In general, if extracted and analyzed properly, the data on social media can lead to useful predictions of certain human related events. Such prediction has great benefits in many realms, such as finance, product marketing and politics, which has attracted increasing number of researchers to this subject. Study of social media also provides insights on social dynamics and public health. A survey provides us perspective and is helpful for carrying out further research.

The rest of the paper is organized as follows: in section 2, some technical backgrounds and basic concepts are introduced. The problems areas where prediction with social media appears promising are described in section 3. Section 4 describes the metrics that are used in prediction. In section 5, we summarize some prediction methods and models. Trends and future work are discussed in section 6. The conclusions are presented in section 7.

## 2. Technical background

In this section, we will introduce basic concepts about social media, and discuss their important characteristics.

**1) Social network**

A social network is a social structure comprising of persons or organizations, which usually are represented as nodes, together with social relations, which correspond to the links among nodes. The social relation could be both explicit, such as kinship and classmates, and implicit, for example friendship and common interest. For instance, fig. 1 is an example of undirected social network in a company, from open source software GUESS [4]. In fig. 1, each node represents an employee. The edge between two nodes means these two employees have some communications in work and the weight of each edge is the communication frequency.

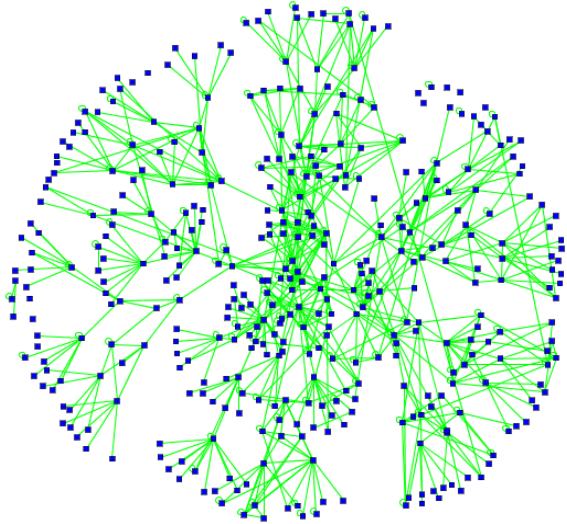

**Fig. 1.** An example of social network from GUESS

A small social network may be modeled by regular graphs such as that of a small world network [5][6][7]. For a large well-connected network, most nodes can reach every other node through a small number of links. The idea of *six degree of separation* suggests that, on average, every two persons are linked by six hops [8]. The situation in online Social Networking Service (SNS) is not much different. The average distance on Facebook in 2008 was 5.28 hops, while in November 2011 it is 4.74 [9]. In the MSN messenger network, which contains 180 million users, the median and the 90th percent degree of separation are 6 and 7.8 respectively [10]. On Twitter, the median, average, and 90th percent distance between any two users are 4, 4.12 and 4.8, respectively [11]. In brief, the degree of separation varies on different SNS platforms and/or on different time but it is quite small.

A social network is a scale free network [12][13] for which the degree distribution asymptotically follows a power law. On Twitter, up to $10^5$ of the number of followings/followers fit the power-law distribution with the exponent of 2.276 [11]. The number of being re-tweeted and mentioned by users on Twitter also follows a power law [14].

**2) Social media**

Social media comprise platforms to create and exchange user-generated content [15][16][17]. Sometimes social media are called consumer-generated media (CGM). Social media are different from traditional media, such as newspaper, books, and television, in that almost anyone can publish and access information inexpensively using social media. In contrast, traditional media (which is also referred as old media or legacy media) requires significant resources to publish contents. But social media and traditional media are not absolutely distinct. For example, major news channels have official accounts on Twitter and Facebook.

There are many forms of social media that include blogs, social networking sites, virtual social worlds, collaborative projects, content communities and virtual game worlds [18]. Some forms of social media lack a social network. Thus in blogspot.com, which is a famous blog platform, there is no social links among bloggers.

Social media has some or all of these seven function blocks: identity, conversations, sharing, presence, relationships, reputation, and groups [19]. Different forms of social media have different points of focus. For example, collaborative projects such as Wikipedia mostly care about sharing and reputation, And in virtual game worlds, identity, presence, reputation, and groups are of the greatest concern.

Recently, social media played important role in unfolding newsworthy events. For example, in the aftermath of the Tohoku Earthquake in Japan people used social media to contact friends, exchange crisis information, and find necessary resources.

**3) Social networking service**

Social networking service is a set of online sites and applications, which at least consist of three parts: users, social links, and interactive communications [20][21]. In fact, SNS is a subset of social media, which include the social network.

On SNS, communication is interactive. For instance, for pure blogs, a non-SNS social media such as blogspot.com, the users' major motivations could be recording one's daily life, providing commentary and opinions, expressing feeling and emotion, demonstrating ideas via text, and keeping community [22]. The first four motivations are all information sharing. For microblogging, a typical SNS, the user intention could be roughly classified into three categories: information sharing, information seeking, and friendship maintenance [23].

All SNS providers have two core focuses: social relations and user-generated contents. In terms of social relations, they might reflect the social network of persons in real life, build new social connections based upon interests and activities, or both. For user-generated contents, they provide an easy way to create, share, rank and exchange information.

**4) Why we need to predict automatically**

Even though currently most predictions using social media can be done better by human agents, specifically experts, there are still good reasons for us to try to predict automatically.

Firstly, compared with human labor, automatic prediction with machines has a much lower cost [24]. Secondly, persons tend to overvalue small probabilities and undervalue high probabilities. So events with small and high probabilities are poorly predicted by people [25]. Thirdly, intentionally or unintentionally, a person may make decision influenced by their desire, interests and benefit, not purely based upon objective probability [25][26]. Lastly, automatic prediction methods could process greater amounts of data and provide response quickly.

# 3. Prediction subjects

In this section, we describe areas where prediction with social media may be made. Generally, a subject, that could be well predictable with social media, must meet the following requirements.

Firstly, the prediction subject must be human related event. On social media, users publish their opinions and beliefs. Prediction methods analyze, extract and integrate the information, and then according to the influence of persons to the predicted subject, make the prediction. But if the subject is non-human-related event, such as eclipse, even though there may be tons of users discuss this topic on social media, the users' thoughts have nothing to do with the development of that event. Consequently, the data on social media could not be used to predict natural events whose development is independent of human actions.

Secondly, if masses of people are involved, the distribution of composition of involved persons on social media should be the same as or similar to that in real world [27]. Because not everyone in real world will use social media, the users on social media could be treated as samples of the involved masses in most cases. But the sampling process is uncontrollable, which may lead to samples with built-in bias. Even though we cannot exclude biased samples completely, we should make sure the proportion of the biased samples is in the acceptable and reasonable range.

Lastly, the involved events should be easy to be talked in public. Otherwise, the contents on social media would be biased [27]. For example, three is social consensus that giving appropriate tips is good and excessively low tipping is impolite and unacceptable. Under such social pressure, almost nobody is willing to admit that he/she paid tips that were too low. The anonymous mode could be used in getting an answer to this issue but such an anonymous mode will have no information about relevant social network structures.

### 1) Marketing

There is some evidence that there is strong correlation between spikes in sale rank and the number of related blog posts. But at the same time, based on blog mentions, predicting whether tomorrow's sales rank for a particular item will be higher or lower than today's sales rank appears to be hard [28]. There are two possible reasons for these seemingly contradictory conclusions. On one hand, there may be a delay between the increase of blog mentions and the increase of sale. On the other hand, the number of blog mentions may predict the change of sale. But the change of sale about one product does not necessarily change the sale rank of other products.

Even though there is correlation, few researchers work on the prediction of sales with social media. Since the daily sales data may involve commercial confidentiality, and there are so many ways to purchase products, it's nearly impossible to get accurate daily sales data. Consequently, researchers prefer to work on a micro level, that is on the product adoption by customers.

Using social media, customers with extremely positive or negative experiences are more likely to express their feelings and evaluation, compared with these with moderate experiences [29]. On Twitter, organization or product brands are referred in about 19% of all the tweets, more than 80% of which do not show any significant sentiment [30]. The mass electronic Word of Mouth (eWOM) gives us a chance to investigate how eWOM and social networking works on product adoption and predict its potential adoption.

The word-of-mouth (WOM) is greatly influential on the first purchase of product or service [31][32], especially when the WOM is from friends or peers [33]. Generally, the negative eWOM is more powerful than the positive ones [34][35][36]. In terms of online recommendations, we find conflicting research results. Some research points out that the more copies of the same message is received, the higher probability that one will adopt that innovation [37]. But in another research, excessive recommendation was seen to have a negative effect. Initially the probability of purchasing increases with more recommendation but after some threshold, the probability drops and stays on a relatively low level [38].

Social networking also affects the production adoption greatly. The possibility of an individual's purchase increases if the product has been strongly adopted by friends [39][40][41]. What is more, users with fewer friends are more easily influenced into adoption [40]. Compared with eWOM, which is one kind of explicit recommendation, social networking influences adoption as kind of implicit recommendation.

The influence of eWOM and social networking on product adoption could be partly explained by Heider's balance theory [42]. In balance theory, friends tend to achieve and maintain consistency in liking and disliking of objects. The consistency will then lead to similar or same product adoption.

The eWOM and social networking do have some impact on product adoption. However, demand comes first in adoption [40]. And individuals could influence just a few friends, rather than everybody they know [38]. So to predict product adoption, we should use social media as an auxiliary prediction tool rather than the decisive one.

**2) Movie box-office**

To predict movie box-office with social media is one of the most studied area. In addition to the traditional prediction factors, such as MPAA rating and number of screens [43][44], social media contents could also be effective to predict box-office [45]. There are many reasons that predicting movie box-office a good subject for research.

Firstly, there are volumes of data about movies and related social media. According to IMDB.com, more than 200 feature films, which originate in the U.S.A and have U.S.A box-office record, were released every year. Besides, movies are widely talked on social media. For example, there are more than 100,000 tweets for each monitored movie [45]. Consequently, there is enough data to be analyzed.

Secondly, the box-office is easy to be accessed and estimated. On one hand, the gross income and opening weekend income is easily obtained from Internet Movie Database (IMDB). On the other hand, the income on opening weekend typically accounts for about 25% of total sales [46]. So we could get the approximate box-office just after the opening weekend. In some cases, the prediction about the high-grossing movies is much accurate than that about low-grossing movies [44]. Even though most researchers treat the box-office as continuous variable, sometimes discretization is applied to divide the box-office into classes according to their amount [43].

Lastly, there is a clear logical correlation between social media contents and movie box-office. The users who post something before the movie release are surely interested in the movie and consequently they are likely to watch the movie. The 1-week pre-release data has the strongest correlation with gross than the data in any other pre-release time periods [44]. After the movie release, user posts, especially the ones with the sentiment [47][48], turn to be kind of eWOM, which would influence other potential customers.

Nevertheless, there are some unique obstacles in research on movie box-office. Usually, the names of movies are also used as other meanings in communications. For example, the very famous movie "The Godfather" by Francis Ford Coppola has a title which would be used in lots of other cases. Actually, it's impossible to find out the really related tweets though search with "The Godfather" as keywords. What is more, some movies hold the same title. For instance, there are 4 movies with the same title of "Love". In such cases, it is quite hard to distinguish the specific movie related social media data and other contents.

Some researchers have tried to predict the Oscars winners using social media [24][48][49]. The voting on all categories of Oscars is restricted to active members of Academy of Motion Picture Arts and Sciences. Although this election process is a professional voting system, and has little to do with the wisdom of crowd, the members of the academy are themselves being influenced by social media. Thus, indirectly, social media is useful in predicting the winner of the Oscars.

**3) Information dissemination**

Information dissemination means how contents spread on the Internet. In other words, it refers to how users pay attention to different information. Since contents attract the users in an asymmetric way, the attention of most users is concentrated on a few contents [50]. Successful prediction could enhance the user experience by providing them with the most attractive information. Information dissemination could be researched on micro level and macro level.

Micro level information dissemination focuses on the adoption of contents between pair of peers [16][51]. To some extent, this issue is similar to the recommendation problem [51]. In the recommendation problem, based upon the previous purchase, the seller tries to find out the possibility that the buyer is interested in buying other items. Paralleling this, in the information adoption problem, the system tries to compute the possibility that a specific user is willing to vote or forward a new post in light of the previous post adoption behavior. Using suitable users' features and content features, we can map the question into recommendation issue and use any online recommendation system and method to predict the post adoption.

Macro level information dissemination illuminates how contents distribute in large scale. The spread depends on both who start it, and how ready the social group is to accept it [16][37]. Compared with micro level prediction, the macro level is much more complicated and there is no universal model that could handle every case. Due to the large volume of diverse samples, the macro level prediction for one specific realm has statistical features and characteristics that are not much influenced by noise.

Different kinds of information have different life cycles [11][52][53]. For example, on Digg.com, the pieces of news get their final popularities in about one day. On Twitter, most topics remain active for a week or less. On YouTube.com, the videos continue to get new views over a very long period. This difference in life cycle length owes to the difference in the posts' inherent value to users [52], which also contributes to the difference in the distribution of posts' adoptions [54]. Prediction on contents with short life cycle is more accurate than that with long life cycle [52].

Furthermore, the effect of social network differs at different stages [52]. At the very beginning of the item's life cycle, when the item is visible to only a few users, the social network could help to promote the post. But after some time, when the information faces lots of persons, the effect of social network is very limited.

Measured by the number of views or votes, early and late popularity is correlated to some extent [52][55]. About half of the re-tweets were brought out within one hour, and 75% within one day from the birth of source tweet [11]. And after logarithmic transformation, there is a strong linear correlation between popularities at early and later, with the residual noise being normally distributed on transformed scale [52].

The macro level pattern varies at different times. In terms of the days in one week, the activities in weekdays are about 50% more than these in weekends [52]. Also in different periods of week, the users' focus varies [23]. For example, the term "school" is used more during the weekdays, while "friends" is more frequent in the weekend. In terms of the hours in one day, users are more active in daytime than in night. On an average, in the first two hours, a post could get about 400 votes if it's posted at 12 pm and only 200 votes if it's posted at 12 am [52].

In sum, different stages in the information life cycle have different suitable predictors. Generally, semantic analysis of contents is more helpful at the very beginning, when no user voting data is available [56]. Then the social network and peer attitude plays an important role in information spreading. Finally with a large user base, prediction could be made based upon the early data.

4) Elections

Election prediction uses the survey of public opinion on political party or politician from a particular sample to predict the election result. Traditionally, the election polls could be done via telephone surveys. But thousands of calls easily lead to cost as high as tens of thousands of dollars. As a newly emerging method, web survey with social media provides an opportunity to do that with low cost.

Simply, the number of related social media contents may be a valid predictor for successful election. In 2008 U.S.A Presidential Primaries, just the number of Facebook supporters could predict the result successfully [57]. In 2009 German federal election, even though 4% of all users are responsible for more than 40% of the contents, the number of messages on Twitter still could have predicted the election result, and it even came close to the tradition election poll's accuracy [58]. The sentiment could also be helpful to do prediction, but not substantially [59].

At the same time, there is an ongoing debate on whether currently social media are effective in collecting public opinions in an unbiased manner and predict the election result. For example, in British Columbia's 2001 provincial election, the number of mentions on internet message boards did not indicate the relative strength of parties [60].

In such research, when comparing the web survey with traditional polls, researchers did not provide the date when the traditional polls were made [61]. Prediction with social media was usually made very close to the election. If the traditional polling was made far from that event, the comparison is unfair and meaningless. This research also excluded small parties [61]. Adding the small parties could have changed the predictions.

The researchers did not point out why a specific time period to collect social media contents was chosen [61]. The prediction result varies heavily depending on the time frame. Including only additional few days leads to a considerable increase in the mean absolute error. No research gives any guideline to choose a reasonable and accurate time window.

The baseline to justify the social media survey method should not always be the random choose [62]. In the U.S.A congressional elections, incumbents won 91.6% of the races in 2008 and 84.5% in 2010. If just indicating all the incumbents would win, the accuracy would be higher than 80%. Using similar methods on different data sets produces worse results than the ones in the original papers [62][63] with the mean average error of 17.1% for using mere Twitter volume, 7.6% for the sentiment analysis, and only 2-3% for traditional professional polling service.

It should be noted that social media does not reflect the demographics of the society. In terms of age, in 2000, 36% of U.S.A citizens between 18 and 24, 50% of citizens between 25 and 34, and 68% of those over 35 voted [62], but on Twitter, more than 60% of users are under 24 [64]. Thus random sampling on social media is biased sampling. What is more, it is hard to know the age of social media users, because the user's profile is confidential. Accordingly, it is near to completely impossible for statistically unbiased sampling on social media, in terms of age, and similarly other attributes, such as region and ethnicity. On the other hand, when applied to political content on social media in general and Twitter in particular, the accuracy of sentiment analysis methods, used in some prediction models, is better than a random classifier to indicate the political orientation of the users [62][63]. One possible explanation is that the vocabulary in most sentiment analysis system is designed for well-written and standard English, rather than the short posts on social media [30][59].

5)  **Macroeconomic**

The macroeconomic includes the regional, national or global economies. Some researchers are trying to use social media to deal with its trends, such as economic indices and stock markets. Generally the social media could not be used to accurately determine these trends alone, but could assist the researchers to capture or predict trends.

In terms of economic indices, researchers have used social media to directly predict or assist in prediction. For Gallup Organization's "Economic Confidence" index and Index of Consumer Sentiment (ICS) from Reuters/University of Michigan Surveys of Consumers, some ratios based upon sentiment information of social media is able to capture the broad trends in traditional economic polls [59]. But the coefficient between sentiment data and consumer confidence varies greatly at different time.

In stock markets, research [66][67] based upon random walk theory and Efficient Market Hypothesis (EMH) suggests that stock prices are unpredictable. But recent research, from the perspective of Socioeconomic Theory of Finance and behavioral economics, suggest that stock prices could be predicted to some extent.

The larger number of postings on finance message boards, such as Yahoo! Finance, predicts negative subsequent stock returns [65]. The correlation between them is statistically significant. But the effect is economically very small but the volume of postings is helpful to predict the stock volatility [65]. For posts that include specific emotive words such as hope, worry and fear, the total number of them is more predictive to stock indices than the number and proportion of their forwarding times and original authors' followers [69]. Similarly, the total number of such words has a strong correlation with other financial market trends, such as gold price, oil price and currency exchange rate [70]. Even though it lacks solid evidence and justification, it appears that a non-linear relationship is likely to exist between social media and stock market. Non-linear models, such as Support Vector Machine (SVM), are able to better utilize social media for prediction [72].

Sentiment information has predictive value. The disagreement of postings, a sentiments measure, can predict the number of trades [65]. Greater disagreement is accompanied with fewer trades on the next day. Additionally, even though general sentiments, such as positive vs. negative, is not helpful in the prediction of qualitative changes in closing values of the Dow Jones Industrial Average (DJIA) and S&P 500, some specific and more detailed sentiments, such as calm, happiness and Anxiety Index, have a predictive power to inform broad direction of stock market in near future [68][71].

Even though some progresses in prediction of macroeconomic with social media have been made, there remain limitations in current research. The prediction effect of social media on the macroeconomic domain in general and stock market in particular appears to be very small. For example, according to [65], a 100% increase in volume of related social media posts will only result in a 0.2% decrease in the stock price. Such a small effect is easily concealed by other factors. Also, existing models lack long term validation [59]. Also, current models do not provide any theory or hypothesis about why and how the social media is able to make macroeconomic predictions. Some researchers treat social media and stock markets as two seemingly unrelated systems [71].

**6) Miscellanea**

In addition to the topics mentioned above, social media is also used for prediction in other realms. Software projects are usually divided into pieces of tasks and finished by many teams simultaneously. Periodically, the works of all teams is aggregated and integrated into the whole project. Since communication influences coordination [73][74], some researchers use user generated comments on tasks during project to construct the communication social network for the whole project, and then combine the social network structure metrics to predict the project integration build failures [75].

Similar to the previous research [75], on a micro level, some other researchers use comments on issues for each file to construct the comment network for every single file, and then integrate the social network analysis metrics to predict the defect of the corresponding file [76].

## 4. The predictors

In this section, we will list the major metrics about social media used in prediction. Mostly these metrics alone do not have sufficient prediction power but their combinations work better. These predictors may be divided into two categories: message characteristics and social network characteristics.

**1) Message characteristics**

Message characteristics focus on the messages themselves, such as the sentiment and time series metrics. If the research focus on general objects, all the available posts are fetched with timestamps. Otherwise, the search result with man-crafted keywords is preferred.

(a) Sentiment metrics

The sentiment metrics are the static features of posts. In addition to the general sentiments discussed in the following, there are some specific sentiment categories, such as happiness and anxiety, on a case-by-case basis. Because they lack generality, we do not investigate them in detail. But the concept, extraction and usage of them are same as these of general ones.

With qualitative sentiment analysis system, the messages could be labeled as positive, negative, or neutral sentiments. Thus naturally the numbers of positive, negative, neutral, non-neutral, and total posts are five elementary content predictors. These metrics may have different prediction power at different stages. For a predicted event in general and movie box-office in particular, the number of positive references correlates with the event result better than total count in the pre-event period. But in the post- event period, the total count is better [47].

Additionally, we could compute the ratios among them, commonly including the ratio between the numbers of positive and total posts [44], the ratio between the numbers of negative and total posts [44], the ratio between the numbers of neutral and total posts, the ratio between the numbers of non-neutral and total posts [44], the ratio between the numbers of neutral and non-neutral posts [45], and the ratio between the numbers of positive and negative posts [45][59]. These ratios reflect the relative strength of these sentiments. More complexly, we can combine these basic elements to compute the sentiments difference [44] and sentiments index $I_{sent}$ [48].

$$Sent\_deff = \frac{N_{positive} - N_{negative}}{N_{total}} \quad (1)$$

$$I_{sent} = 100 * \left( \frac{\frac{N_{positive} - N_{negative}}{N_{total}}}{2} + 0.5 \right) \quad (2)$$

The $N_{positive}$, $N_{negative}$ and $N_{total}$ mean the number of positive post, negative posts and total posts respectively. The sentiments index is proved to have strong correlation with IMDB rating and be useful in prediction Oscar prizes when used for movies.

(b) Time series metrics

Time series metrics try to investigate the posts dynamically, including the speed and process of the message generation. The posts generating rate means how quickly the messages are produced. It's easily computed as the following:

$$Post\_rate = \frac{N_{total}}{Time\ window\ size} \quad (3)$$

According the time windows size, the generating rate could be estimated different, such as hourly, daily or weekly. With higher posts generating rate, more persons are concerning it, and the topic is more attractive. Some experiments showed that, the daily generating rate before release is a good predictor for movie box-office [45].

The source composition of messages in a time window is another time series metrics. Here the time could be real or virtual. For example, in voting on digg.com, we could treat each vote as a Digg second. So in the first ten Digg seconds, the votes are composed of fan votes and non-fan votes. If the fan votes occupy overwhelming proportion of the whole, these posts will finally accumulate fewer votes than others [54]. Because for each interval between two fan votes, the more non-fan votes are in the interval, which indicates the post's attraction to public, the less fan votes' proportion is, and the more the final votes will be.

**2) Social network characteristics**

Social network characteristics measure structure features. We also call these characteristics as metrics/measures in social network analysis. Being long studied, these are so many characteristics that it's impossible to list and investigate all of them here. So we just enumerate and briefly discuss the most used ones in predictions.

(a) Terminology

Here we introduce some special terms used in social media. Since different social networking sites have different functions and names for the social connections, we unify the namespace to simplify the discussions. In the following, we will focus on the directed network. And the undirected social network is similar. Unless specified, all the discussions are with directed networks.

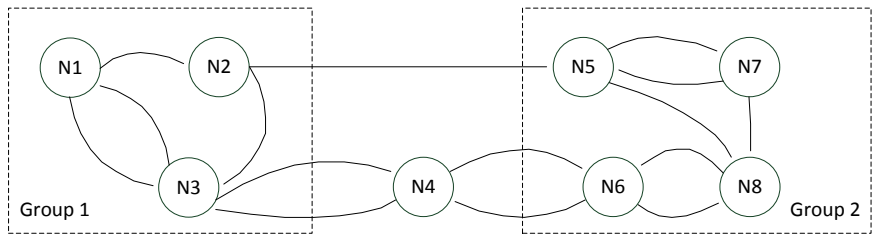

**Fig. 2.** An example of a directed social network

**Node**: every node represents one unique entity in the social networking. For example, on Twitter, the users could be expressed as nodes. While on an enterprise trading network, the companies are denoted as nodes.

**Follow**: if node A indicates to have a relationship with node B, A follows B, which is represented as a directed line from A to B in graph. For example, in the Fig. 2, N1 follows N2. The relationship could means differently in different sites. On YouTube, it is subscription. While on Twitter, it is being fans. Additionally, the follow could be unidirectional or bidirectional. In unidirectional follow, such as Google+, node A could follow node B, without being followed by B. Oppositely in bidirectional mode, such as Facebook, the follow from node A to node B will accompany with the follow from B to A.

**Follower**: if A follows B, A is B's follower. For instance, in the Fig. 2, N5 has N2 being its follower.

**Followee**: followees are the following entities. In Fig. 2, N5 and N3 are N2's followees. And N1 is the follower and followee of N3 at the same time.

(b) Degree

Degree is the number of ties to/from other nodes in the network, including in-degree, out-degree and total-degree. The in-degree and out-degree are available only when the social network is directed. Sometimes, the degree of a node is also named as its degree centrality.

**In-degree**: in-degree of a node is the number of directed lines to this node. That is, the in-degree is the count of followers. On average, each user has 85 followers on Twitter [77]. Treating the count of posts as the activity index, with the in-degree increasing, the activity firstly increases, and then after about 300 followers as threshold, become stable [77].

**Out-degree**: out-degree of a node is the number of its tail endpoints. In other words, the out-degree is equal to the number of followees. Averagely, each Twitter user follows 80 other users [77]. Because the research [77] only samples a fraction of Twitter users rather than all of them, the average values of followers and followees are not the same. Both the in-degree and out-degree alone indicate litter about the user's influence [14][40].

**Total-degree**: total-degree is the sum of the in-degree and out-degree. In the undirected network, this is the only one degree metric.

(c) Density

**Density**: density is the proportion of existing edges count relative to maximum possible edges count. In a directed social networking with *n* nodes and *e* edges, the density is computed as:

$$Density = \frac{e}{n(n-1)} \tag{4}$$

And in an undirected social networking with same parameters, the density is estimated as:

$$Density = \frac{e}{n(n-1)/2} \tag{5}$$

The network of followers/followees is very dense. But in 2009, 77.9% user pairs have one-way connection. That is, only 22.1% user pairs have reciprocal connections [11]. Thus the network of actual friends, who communicate reciprocally and directly, is much sparser and simpler [11][77].

(d) Centrality

Centrality measures the relative importance of a node within a network. Betweenness centrality and closeness centrality are two centrality metrics, which are widely used. Furthermore, in addition to the degree centrality on the node level, which is discussed in section (b), the group degree centrality is introduced.

**Betweenness centrality** [78]: the betweenness centrality quantitatively measures the control of a node on the communication between other nodes in the social networking. In a network with n nodes set *V*, the betweenness centrality of a node *v* is:

$$C_B(v) = \sum_{s \neq v \neq t \in V} \frac{\delta_{st}(v)}{\delta_{st}} \tag{6}$$

where $\delta_{st}$ is the total number of shortest paths between node *s* and node *t* and $\delta_{st}(v)$ is the number of these paths passing through *v*. The betweenness centrality could be normalized by $n(n-1)$ for directed network or $n(n-1)/2$ for undirected network.

**Closeness centrality** [79]: in a network, the distance of two nodes is the length of the shortest path between them. The farness of a node is the sum of its distances to all other nodes. And its closeness centrality is the inverse of the farness:

$$C_C(v) = \frac{1}{\sum_{s \neq v \in V} lenght(SP_{sv})} \tag{7}$$

where $SP_{sv}$ means the shortest path between node *s* and *v*. For node *v*, the closeness centrality estimates how long it will take to transmit information from node *v* to all other nodes in the network.

**Group degree centrality** [80]: extent the degree from the node level to group level, the degree centrality of a group *G* with *n* nodes is introduced as following:

$$C_D(G) = \frac{\sum_{c_i \in G}(C_D(c^*) - C_D(c_i))}{(n-1)(n-2)} \tag{8}$$

where $C_D(c_i)$ is the degree of node $c_i$ and $c^*$ is the node with the highest degree in *G*.

(e) Structural hole

The structural hole theory measures the positional status of each node in its ego network. The ego network of node *i* is a sub-network of the whole, which consists of nodes *i* (ego) and all neighbors of node *i* [81]. In the following, we will introduce two numerical metrics in structural hole theory.

**Effective size**: for node *v* with its *n* nodes ego network *EN*, the effective size is computed as:

$$ES(v) = n - 1 - \frac{\sum_{c_i \neq v \in EN} Degree(c_i) - Degree(v)}{n-1} \tag{9}$$

That is, the effective size of node *v* equals to the number of its neighbors minus the average degree of neighbors, not including their connections to node *v*. For example, in Fig. 2, N2's ego network includes N1, N2, N3 and N5. And the effective size of N2 is 3-4/3 = 5/3.

**Efficiency**: the effective size normalized by the number of neighbors:

$$E(v) = 1 - \frac{\sum_{c_i \neq v \in EN} Degree(c_i) - Degree(v)}{(n-1)^2} \tag{10}$$

In addition to the metrics mentioned about, there are still lots of other social media characteristic, such as network diameter. But they are not widely used or proved to be powerful in prediction. So we do not list them in detailed here. Besides, our task is how to use these metrics to predict. Thus in the previous part of this section, we just list and briefly discuss them, without deeper insight into them.

## 5. Prediction methods

In this section, we discuss some methods used in prediction with social media.

**Regression method**: Regression methods analyze relationship between the dependent variable, prediction result, and one or more independent variables, such as the social network characteristics. Regression model could be linear and non-linear. But the linear model seems to describe the relation best [48]. Thus most times, we use the linear regression models, rather than non-linear ones, such as exponential, logarithmic, and polynomial models. In linear regression model, the variables could be the raw or transformed data. For example, between the early and late popularities of posts on digg.com, the correlation is based upon the logarithmically transformed data [52]. Besides, sentiment data does not work well in the regression models for movies [44]. Currently, this is the simplest and most used method.

**Bayes classifier**: Bayes classifier is a probabilistic classifier using Bayes' theorem. Based upon the priori probability of the prediction event, Bayes classifier uses the Bayesian formula to calculate its posterior probability, that the object belongs to the result classes, and then select the class with the largest posterior probability, as the event is most likely to have that result. If the prediction result is discrete, the Bayes classifier can be applied directly. Otherwise, the prediction result must be discretized first [43]. This classifier has an assumption that the predictors must be conditionally independent. There is no solid evidence always that the discussed metrics satisfy this assumption.

**K-nearest neighbor classifier**: K-nearest neighbor classifier, one of the simplest machine learning algorithms, tries to cluster the objects according to their distance to others. Commonly we use the Euclidean distance and Manhattan distance. For two entities $p = \{p_1, p_2, ..., p_n\}$ and $q = \{q_1, q_2, ..., q_n\}$ with n-dimensional feature vector, the Euclidean distance is computed as:

$$ED(p,q) = ED(q,p) = \sqrt{\sum_{i=1}^{n}(p_i - q_i)^2} \qquad (11)$$

The Manhattan distance is calculated as:

$$MD(p,q) = MD(q,p) = \sum_{i=1}^{n}|p_i - q_i| \qquad (12)$$

Then, the entity, which is being predicted, is assigned to the cluster including most of its k-nearest neighbors. Finally, the entity is predicted to have the same output as the entities in its cluster.

**Artificial Neural network**: Artificial neural network is a computational model [82]-[88] to simulate the human brain. An artificial neural network consists of lots of artificial neurons. And these neurons could belong to many interconnected group, including input layer, hidden layer and output layer. The input layer is responsible for receiving raw data and transmitting them to the next layer. The output layer will give us the final prediction result. Thus using artificial neural network to do prediction, our major task is choosing the network structure and designing the hidden layer. In addition to using them to predict directly, the Self Organizing Map (SOM), one kind of artificial neural network, could be used for features' dimensionality reduction for further analysis.

**Decision tree**: Decision tree is a visual technique in data mining and machine learning. Travelling from root node to leaf, one entity will get the prediction result. Classification tree and regression tree are two basic and major types of decision trees. Classification tree analysis is applied when the prediction output is discrete classes. And regression tree is used when predicted outcome is continuous value. Unlike the artificial neural network being a black box model, the decision tree is a white box model, which could be relatively easily explained. Besides, decision tree works well with dummy variables and empty variables.

**Model based prediction**: This possibly is the hardest way to do prediction. We have to build a mathematical model on the object before prediction, which requires deep insight into the object. At this point we do not know enough about social media to develop effective models for them. Even though there is some progress in modeling [16][54], model-based prediction remains an open and challenging topic.

## 6. Future work

As an emerging research topic, prediction with social media faces many challenges. Here we point out some urgent and important future works.

**Using sociological theory to interpret predictors**: Currently, researchers choose predictors using the trial and error method. We know neither why these predictors are better than others, nor how these predictors could predict the result. Not knowing the background logic between these metrics and the final prediction result, we just use a collection of metrics to be trained on test data, find out which ones have the highest coefficients, and use them to compose the prediction model. Consequently, lacking a solid supporting theory, we cannot be sure that one model, which works well in one case, could be applied to other situations with the same accuracy. That's why some models show good performance in one election prediction, but completely fail in another one [58][61]. To guarantee our model has good performance in all cases, we need to know the logic and theory behind the model.

**Trying more prediction methods**: Most researchers use simple methods such as linear regression analysis. These methods are known to work well under some conditions. Social media is produced on a complex system and thus more likely than not the predictors and prediction outcomes have non-linear correlation. Furthermore, combination of methods might lead to breakthrough [82]. In such combination, a surface learning agent, such as instantaneously trained neural networks, quickly adapts to new modes and emerging trends on social media. And a deep learning agent focuses on long-term patterns. In a nutshell, we should try some non-linear methods and find out the suitable methods and/or combinations for each prediction realms.

**Modeling on predictions with social media**: We are far from knowing everything about social media. For instance, there are different kinds of prediction objects which show different features. Taking recommendation adoption as an example, the recommendation on DVDs is more likely to be accepted than that on books [38]. But there is still no universal accepted conclusion about why these differences exist. This lack of understanding adds to the difiisulties of modeling. Formal modeling could be necessary and helpful to understand and investigate the features and behaviors of prediction techniques.

**Semantic analysis system for social media**: Although semantic analysis is not a necessary part of the prediction methods, it is frequently used. Thus the accuracy of semantic analysis is critical to the prediction performance. Semantic analysis could be based upon lexicon or previous statistics. In terms of lexicon, compared with natural and formal English language, social media content has similar structure, but many different words [30], such as "lol", which short for "laughing out loud". This Internet slang affect the semantic analysis system, because the lexicon in most existing systems is designed for well-written English [59]. Besides, Internet slang evolves quickly and chaotically. The SOM, which sometimes is used to construct thesaurus as an unsupervised or semi-supervised clustering method, could be helpful in this issue. These methods firstly label some posts manually and then use statistical model, such as naïve Bayes classifier, to mark other posts according to statistical features of labeled ones. In some forms of social media, such as microblogging, the length of post is so short that it shows no significant statistical characteristics.

## 7. Conclusions

In this paper, we presented a survey of prediction using social media. We also gave an overview of prediction factors and methods and listed challenging problems and areas for further research. Although prediction using social media is only an emerging research topic and its results have relatively low accuracy, it has created a new way for us to collect, extract and utilize the wisdom of crowds in an objective manner with low cost and high efficiency.